\documentclass[12pt]{article}
\let\section=\subsection     \let\subsection=\subsubsection                
\usepackage{graphicx}
\usepackage{rotating}
\newcommand{\be}{\begin{eqnarray}}
\newcommand{\ee}{\end{eqnarray}}
\newcommand{\non}{\nonumber\\}
\newcommand{\raf}[1]{(\ref{#1})}
\newcommand{\ave}[1]{\left\langle #1 \right\rangle}
\newcommand{\ebe}{E-by-E\ }

\begin{document}
\begin{center}
   {\large \bf Event-by-event fluctuations in heavy ion collisions}\\[2mm]
   M.~D\"oring and V.~Koch\\[5mm]
   {\small \it  Gesellschaft f\"ur Schwerionenforschung (GSI) \\
   Postfach 110552, D-64220 Darmstadt, Germany \\
   and\\
   Lawrence Berkeley National Laboratory\\
   Berkeley, CA, 94720, U.S.A. \\[8mm] }
\end{center}

\begin{abstract}
  We discuss the physics underlying event-by-event fluctuations in
  relativistic heavy ion collisions.
  We will argue that the fluctuations of the ratio of
  positively over negatively charged particles may serve as a unique
  signature for the Quark Gluon Plasma.
\end{abstract}

\section{INTRODUCTION}
Any physical quantity measured in experiment is subject to fluctuations.
In general, these fluctuations depend on the properties of the system
under study (in the case at hand, on the properties of a fireball created 
in a heavy ion collision) and may contain important information about the 
system.

The original motivation for event-by-event (\ebe) studies in ultra relativistic
heavy ion collisions has been to find indications for 
distinct event classes. In
particular it was hoped that one would find events which would carry the
signature of the Quark Gluon Plasma. First pioneering experiments in this
direction have been carried out by the NA49 collaboration 
\cite{NA49}. They analyzed the \ebe fluctuations of the mean transverse
momentum as well as the kaon to pion ratio. 
Both observables, however, did not
show any indication for two or more distinct event classes. Moreover, the
observed fluctuations in both cases were consistent with pure statistical
fluctuations. 

On the theoretical side, the subject of \ebe fluctuations has recently gained
considerable interest. Several methods to distinguish between statistical and
dynamic fluctuations have been devised \cite{mrow,voloshin}.
Furthermore the influence of hadronic resonances and
possible phase transitions has been investigated
\cite{shuryak,jeon1,BH99,heiselberg_jackson,heiselberg_report}. All these
theoretical considerations assume that the observed fluctuations will be
Gaussian and thus the physics information will be in the width of the
Gaussian, which is controlled by 2-particle correlations \cite{bialas}.

\section{EVENT-BY-EVENT FLUCTUATIONS} 
Fluctuations have contributions of different nature.
Besides the statistical fluctuations due to a finite 
number of particles  in case of heavy ion collisions, there are also 
fluctuations of the volume. Both these fluctuations are trivial and add to the
dynamical fluctuations which carry the real information about the properties of
the system. The dynamical fluctuations are controlled by the appropriate
susceptibilities,  which are the second derivative of the free energy
with respect to the appropriate conjugate variable. 
For example the fluctuations of the
charge are given by
\be
\ave{(\delta Q)^2} = - T \frac{\partial^2 F}{\partial \mu_Q^2} = -V T \chi_Q
\label{charge_suscept}
\ee
where $\mu_Q$ is the charge chemical potential, $T$ the temperature and $V$
the volume of the system. $\chi_Q$ is the charge susceptibility. 
It is interesting to note that the very same susceptibility also controls the
response of the system to an external electric field. 
The properties of a macroscopic system are studied by investigating 
its response, i.e. its
susceptibility, to external forces. This is of course impossible for the 
mesoscopic systems created in a heavy ion collision. However, the same
property can also be accessed via fluctuations. 

Since we will mostly concentrate on the charge susceptibility, 
let us point out that they are directly related to the electric mass, which is
given by the zero momentum limit
of the static current-current correlation function \cite{Kapusta} $
\Pi_{00}(\omega=0,\vec{k}\rightarrow 0)$.
\be
\ave{(\delta Q)^2} = - V T \Pi_{00}(\omega=0,\vec{k}\rightarrow 0)
\ee
This object has been, evaluated in Lattice QCD \cite{Lattice,gupta} as well 
as in effective hadronic models (see below and \cite{Kapusta,Song,Raju}). 
Note also that this
is the same current-current correlation function, which controls dilepton and
photon production in heavy ion collisions \cite{Koch_qm,wambach}.

\subsection{Fluctuations of particle ratios}
\label{rat_fluct}
In order to avoid volume fluctuations one needs to study observables which are
independent of the volume of the system. Among others the ratio of particle
multiplicities will have this property.
This is certainly true if one
looks at similar particles such as $\pi^+$ and $\pi^-$, where the freeze out
volumes are expected to be the same. Some residual volume fluctuations may be
present if one considers ratios of particles with different quantum numbers
such as the $K/\pi$ ratio, but they still should be small.
Let us define the particle ratio $R_{12}$ of two particle species $N_1$ and $N_2$
\be
R_{12} = \frac{N_1}{N_2}
\ee
The fluctuations of this ratio are then given by \cite{mrow,jeon1,BH99}
\be
\frac{(\delta R_{12})^2}{\ave{R_{12}}^2}  
& = &
\left(\frac{\ave{(\delta N_{1})^2}}{\ave{N_{1}}^{2}} + 
\frac{\ave{(\delta N_{2})^2}}{\ave{N_{2}}^{2}} 
\right. 
-2
\left. 
\frac{\ave{\delta N_{1} \delta N_{2}}}{\ave{N_{1}}\ave{N_{2}}} \right) .
\label{eq:ratio-fluct}
\ee 
The last
term in Eq.~(\ref{eq:ratio-fluct}) takes into account correlations between the
particles of type 1 and type 2. This term will be important if both particle
types originate from the decay of one and the same resonance. For example, in
case of the $\pi^+/\pi^-$ratio, the $\rho_0$, $\omega$ etc.
contribute to these correlations. 
Also this term is responsible to cancel out all volume fluctuations
\cite{jeon1}.

Let us note that the effect of the
correlations introduced by the resonances should be most visible
when \( \ave{N_{1}}\simeq \ave{N_{2}} \).
On the other hand, 
when \( \ave{N_{2}}\, \gg \, \ave{N_{1}} \),
as in the $K / \pi$ ratio, 
the fluctuations are dominated by the less abundant particle 
type and the resonances feeding into it. The correlations are then very hard
to extract. In \cite{jeon1} it was shown that in case of the $K/\pi$-ratio 
resonances and quantum statistics
give rise to deviations from the statistical value of at most 2~\%, in
agreement with experiment \cite{NA49}.

As pointed out in \cite{jeon1} the measurement of particle ratio fluctuations
can provide important information about the abundance of resonances at chemical
freeze out, and thus provides a crucial test for the picture emerging from the
systematics of single particle yields \cite{stachel}. In particular the
fluctuations of the $\pi^+/\pi^-$-ratio should be reduced by about 30 \% as
compared to pure statistics due to the presence of hadronic resonances with
decay channels into a $\pi^+$-$\pi^-$-pair at chemical freeze out.  About 50~\%
of the correlations originate from the decay of the $\rho_0$ and the $\omega$
mesons. Thus the fluctuations provide a complementary measurement to the
dileptons. 

\section{Fluctuations and correlations}
As already pointed out in the previous section, the fluctuations are sensitive
to correlations among the particles of concern. In case of charge fluctuations
which we will discuss below, the relation between the charge fluctuations and
the 2-particle correlation function $C_2$ for a system of unit charged
particles is given by \cite{bialas_02,jeon_pratt}
\be
\ave{(\delta Q)^2} = C_2(+,+) + C_2(-,-) - 2C_2(+,-) + \rho_1(+) + \rho_1(-)
\ee
where the correlation function $C_2$ is defined as
\be
C_2(\pm,\pm) = \int_{\Delta y} dy_{1,\pm} dy_{2,\pm} \frac{d^2N}{dy_{1,\pm} 
dy_{2,\pm}} -
\frac{dN}{dy_{1,\pm}} \frac{dN}{dy_{2,\pm}} 
\ee
The single particle density $\rho_1(\pm)$ is defined as
\be
\rho_1(\pm) = \int_{\Delta y} dy_{\pm} \frac{dN}{dy_{\pm}}
\ee
Here $\Delta y$ denotes the range of acceptance e.g. in rapidity, over which
the particles are measured.
This expression can easily be extended to a system of particles with charge
states different from unity \cite{bialas_02}.

Thus, an alternative way to access charge fluctuations is the measurement of
the one and two particle densities. However in this case the effect of volume
fluctuations still needs to be removed. This can be achieved by defining
appropriate modifications of the correlation functions as discussed in
\cite{voloshin_2}. Also the so-called balance functions introduced in
ref. \cite{balance} can be viewed as another way of 
analyzing these correlation
functions and a relation between the ratio fluctuations discussed here has
been made in \cite{jeon_pratt}.

\section{Charge fluctuations}
 Measuring the charge fluctuations or more precisely the charge fluctuations
 per unit degree of freedom of the system created in a heavy ion collision
 would tell us immediately if we have created a system of quarks and gluons
 \cite{jeon2} (see also \cite{mueller}).
 The point is that in the QGP phase, the unit of charge is $1/3$
 while in the hadronic phase, the unit of charge is 1.
 The net charge, of course does not depend on such subtleties, but the
 fluctuation in the net charge depends on the {\em squares} of the 
 charges and hence are strongly dependent on which phase it originates
 from.  However, as discussed in the previous section, 
 measuring the charge fluctuation itself is plagued by
 systematic uncertainties such as volume fluctuations, which can be avoided if
 one considers ratio fluctuations. 
 The task is then to find a suitable ratio whose fluctuation 
 is easy to measure and simply related to the net charge fluctuation. 

 As shown in \cite{jeon2} the fluctuation of the ratio  $R = N_+/N_-$ 
 serves this purpose and the observable to investigate is  
 \be 
 D \equiv \ave{N_{\rm ch}}\ave{\delta R^2} = 4\ave{N_{\rm ch}}\ave{\delta F^2} 
 =4 {\ave{\delta Q^2} \over \ave{N_{\rm ch}}}
 \ee
 which provides a measure of the charge fluctuations per unit entropy.

 In order to see how this observable differs between a hadronic system and a
 QGP let us compare the value for $D$ in a pion gas and in a simple model
 of free quarks and gluons. Ignoring small corrections due to quantum
 statistics a simple calculation gives for a pion gas 
 \be
 D_{\rm \pi-gas} \approx 4
 \;.
 \label{eq:pion_gas}
 \ee
 In the presence of resonances, this value gets reduced by about 30 \% due to 
 the
 correlations introduced by the resonances, as discussed in section 
 \raf{rat_fluct}. 

 For a thermal system of free quarks and gluons we have in the absence of
 correlations and ignoring small correction due to quantum statistics 
 \cite{jeon2}.
 \be
 D_{\rm QGP} \simeq 3/4.
 \ee
 
 Actually the charge fluctuations $\ave{(\delta Q)^2}$ have been evaluated in
 lattice QCD along with the entropy density \cite{Lattice}. Using these
 results one finds
 \be
 D_{\rm Lattice-QCD} \simeq 1 - 1.5
 \ee
 where the uncertainty results from the uncertainty of relating the entropy
 to the number of charged particles in the final state. Actually the most 
 recent  lattice result \cite{gupta} for the charge fluctuations, 
 which was obtained in the quenched approximation, is somewhat lower then 
 the result of \cite{Lattice}.

 But even using the larger value of $D = 1.5$ for the Quark Gluon Plasma,
 there is still a factor of 2 difference between a hadronic gas and the QGP,
 which should be measurable in experiment.

 The key question of course is, how can these reduced fluctuations survive the
 hadronic phase. This has been addressed in
 \cite{jeon2} and in more detail in \cite{stephanov}. The essential point why
 this signal should survive is that charge in conserved which leads to a very
 slow relaxation of the initial fluctuations.

 Finally one needs to take into account that the total charge of the system in
 conserved and thus does not fluctuate. This has been addressed in
 \cite{bleicher} and a corrected observable $\tilde{D}$ has been derived.

 In Fig. \raf{fig:bleicher} the importance of these corrections, in
 particular the charge conservation correction is demonstrated. There we
 compare the uncorrected observable $D$ with the corrected observable
 $\tilde{D}$ as a function of the width of the rapidity window based on a
 URQMD \cite{URQMD} simulation. The effect of
 the charge conservation is clearly visible. With increasing rapidity window
 the charge fluctuation are suppressed. Once the charge-conservation
 corrections are applied, the value for $\tilde{D}$ remains constant at
 the value for a hadron gas of $\tilde{D} \simeq 3$. This is to be expected
 for the URQMD model, which is of hadronic nature and does not have
 quark- gluon degrees of freedom.

\begin{figure}[htb]
\begin{picture}(100,180)
\put(70,-15){
\includegraphics[width=9cm]{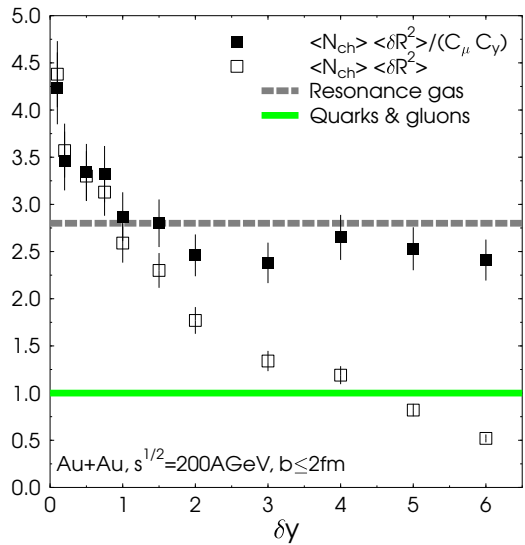}}
\end{picture}
\caption[]{Charge fluctuations as a function of the size of the rapidity window.}
\label{fig:bleicher}
\end{figure}

Also for small $\Delta y$ the value is  $\tilde{D} \simeq 4$
before it drops to $\tilde{D} \simeq 3$ for $\Delta y > 1.5$. This
effect, which was already predicted in \cite{jeon1}, is simply due to the fact
that the correlation introduced by the resonances gets lost if the
acceptance window becomes too small.

Recently an interesting comparison \cite{gale} between predictions for
$\tilde{D}$ by different event generators has been made. The authors find that
indeed a parton cascade model leads to small values for $\tilde{D}$ provided
hadronic rescattering is turned off. This is in line with the predictions of
\cite{jeon2}. However, the authors also report substantial corrections due to
hadronic rescattering, which appear quite large given the arguments presented
in \cite{jeon2,stephanov}. But none of the event generator gave result large
than $\tilde{D} \simeq 3$.

\section{Effects of particle interactions}
\begin{figure}[htb]
\begin{picture}(200,150)
\put(60,0){
\includegraphics[width=9cm]{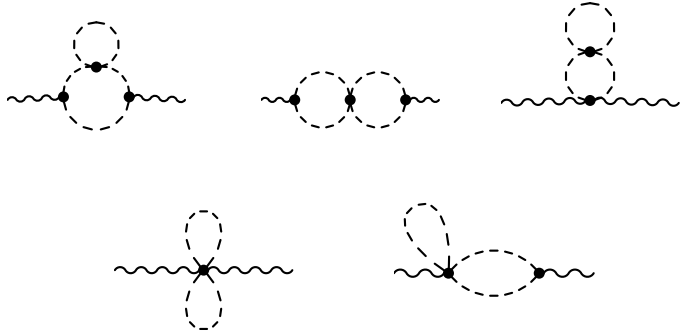}}
\end{picture}
\caption[]{Two loop diagrams contributing to $\Pi_{00}$ to order $g^2$.}
\label{fig:diagrams}
\end{figure}

So far we have discussed the fluctuations of free, noninteracting particles
only. Of course, by considering resonances as well, some of the interactions
are taken into account \cite{Raju_2,Raju}. To address the effect of
interactions we have calculated the electromagnetic current-current correlator
$\Pi^{00}$ as well
as the entropy to the first order of the interaction for a system of pions
interacting via $\rho$-meson exchange. For simplicity, we have assumed that
the $\rho$-mass is infinite,  which leads to point-like interaction among the
pions, and is goverened by the following effective Lagrangian
\be
{\cal L}&=&-\frac{1}{4} F_{\mu\nu}^2-m_\pi^2|\phi|^2
+ \frac{1}{2}( (\partial_\mu \Phi_0)^2 - m_\pi^2 \Phi_0^2)
+|D_\mu\phi|^2
-\frac{g^2}{2
  m_\rho^2}\big(\phi^\star\stackrel{\leftrightarrow}{D_\mu}\phi\big)^2 \non
&&
+ \frac{g^2}{m_\rho^2}
\left(\phi_0 D_\mu \phi-\phi
\partial_\mu \phi_0\right) \left( \phi_0 \left( D^\mu
\phi\right)^\star-\phi^\star\partial^\mu
\phi_0\right)
\label{effL}
\ee
The 2-loop diagrams for the corrections to $\Pi_{00}$ are shown in
fig \ref{fig:diagrams}.

\begin{figure}[htb]
\begin{picture}(200,170)
\put(100,10){
\includegraphics[width=6cm]{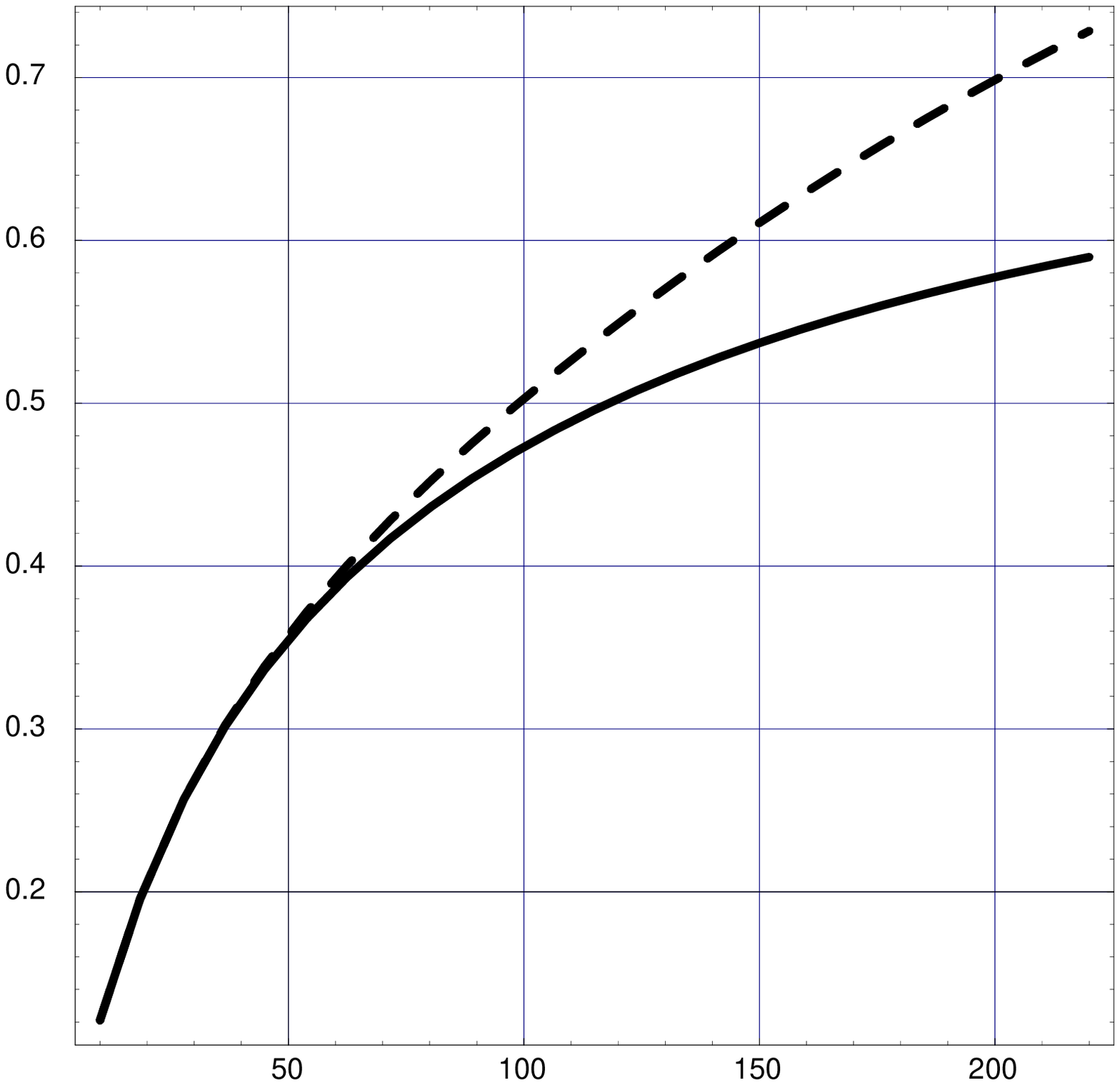}}
\put(140,-3){{\footnotesize Temperature [MeV]}}
\put(70,70){\begin{turn}{90}
$\displaystyle{{\scriptstyle T \;\frac{\Pi^{00}(q_0=0,{\bf q}\to {\bf 0})}{S(T)}}}$
\end{turn}}
\end{picture}
\caption[]{Ratio of charge fluctuations over entropy for free particles
(full line) and with  interactions (dashed line).}
\label{fig:results}
\end{figure}

The ratio $\Pi_{00}(T) / S(T)$ for the entropy $S(T)$ and the self energy $\Pi_{00}(T)$ is shown in
fig \ref{fig:results}.

The effect of the interactions is to {\em increase} the charge fluctuations
per entropy. At temperature of $\sim 140 \, \rm MeV$ the corrections are quite
large, in line with the findings of \cite{Raju}, where chiral perturbations
theory has been used. However, in the same paper \cite{Raju} the effect the
charge fluctuations have also been evaluated in a virial expansion, leading to
much smaller (but still positive) corrections. Therefore, it appears that
 at least dynamical $\rho$ mesons need to be taken into account before any
 quantitative conclusion can be made.

\section{Conclusions}
We have discussed event-by-event fluctuations  in heavy ion
collisions. These fluctuations may provide useful information
about the properties of the matter created in these collisions, as long as the
`trivial' volume fluctuations, inherent to heavy ion collisions, can be
removed. We have argued that the fluctuations of particle ratios is not
affected by volume fluctuations.  

In particular the fluctuations of the ratio of positively over negatively
charged particles measures the charge fluctuation per degree of freedom. Due
to the fractional charge of the quarks, these are smaller in a QGP than in a
hadronic system. 

A measurement of our observable $\tilde{D} \simeq 1$ 
would provide strong evidence for the existence of a QGP in these
collisions. A measurement of $\tilde{D} \simeq 3$ on the other hand does not
rule out the creation of a QGP. There are a number of caveats (see
\cite{jeon2}), which could destroy the signal, such as unexpected large
rapidity shifts during hadronization. 

At this workshop first preliminary results for the results 
observable $\tilde{D}$ have been reported by the STAR, CERES and NA49
collaboration \cite{experiments}. They all find a value which
 is consistent with  $\tilde{D} \simeq 4$. This finding is somewhat surprising since not even the
 effect of resonances seem to be visible in the fluctuations. At the same time
 STAR reports that the measured number of $K^*$ mesons is well in line with 
the expectations of the thermal model \cite{harris,redlich}. So were are the
correlations? Could it be that the interactions among the resonance are indeed
so strong that they entirely compensate the reduction of  $\tilde{D}$ due to
resonances. Also, as already mentioned, in the study presented in  \cite{gale}
{\em none} of the event generators finds a value of $\tilde{D} > 3$ for
rapidity windows $\Delta y > 2$. This certainly needs further investigation. 

Finally let us note that fluctuations of the baryon number in principle can
also be utilized, since in the QGP quarks carry fractional baryon
number \cite{mueller}. This, however, would require the measurement 
of neutrons on an event by event basis.

\section*{Acknowledgments}
This work was supported GSI, Darmstadt and 
by the Director, Office of Energy Research,
Office of High Energy and Nuclear Physics, Division of Nuclear Physics,
the Office of Basic Energy
Science, Division of Nuclear Science, of the U.S. Department of Energy under
Contract No. DE-AC03-76SF00098. V.K. would like to thank GSI for the 
hospitality during his sabbatical leave, where some of this work has been
carried out.

\end{document}